\documentclass{article}
\usepackage[a4paper]{geometry}
\usepackage[utf8]{inputenc} 
\usepackage[T1]{fontenc}    
\usepackage{booktabs}       
\usepackage{amsfonts}       
\usepackage{lipsum}
\usepackage{graphicx}
\usepackage{subfiles}
\graphicspath{ {./images/} }
\usepackage{amsmath}
\usepackage{mathtools,multicol}
\usepackage{caption}
\usepackage{subcaption}
\usepackage{multirow}
\usepackage{authblk}
\usepackage{lastpage}
\usepackage{graphicx, wrapfig, subcaption, setspace, booktabs}
\usepackage[T1]{fontenc}
\usepackage[font=small, labelfont=bf]{caption}
\usepackage{fourier}
\usepackage[protrusion=true, expansion=true]{microtype}
\usepackage[english]{babel}
\usepackage{sectsty}
\usepackage{amsthm}
\usepackage[numbers,sort]{natbib}
\usepackage{url, lipsum}
\usepackage{enumerate}
\usepackage{bigints}
\usepackage[utf8]{inputenc}
\usepackage{hyperref}
\hypersetup{colorlinks}
\usepackage{appendix}
\usepackage{float}
\usepackage[table]{xcolor}
\usepackage{comment}
\usepackage{booktabs}
\usepackage{pifont}
\definecolor{forestgreen}{rgb}{0.0, 0.5, 0.0}

\newtheorem{proposition}{Proposition}

\newtheorem{assumption}{Assumption}

\newtheorem{corollary}{Corollary}
\newtheorem{lemma}{Lemma}

\usepackage[noabbrev, capitalise]{cleveref}
\crefname{axiom}{Axiom}{Axioms}
\crefname{assumption}{Assumption}{Assumptions}
\crefname{lemma}{Lemma}{Lemmas}
\crefname{proposition}{Proposition}{Propositions}
\crefname{theorem}{Theorem}{Theorems}
\crefname{table}{Table}{Tables}
\crefname{equation}{Equation}{Equations}
\crefname{hypothesis}{Hypothesis}{Hypotheses}

\newcommand{\R}{\mathbb{R}}
\newcommand{\reals}[1]{\R^{#1}}
\newcommand{\matsq}[1]{\mathcal{M}_{#1}(\R)}

\newcommand{\mat}[1]{\mathcal{M}_{#1}(\R)}

\newcommand{\orth}[1]{\mathcal{O}_{#1}(\R)}
\newcommand{\spd}[1]{\mathcal{S}_{#1}(\R)}
\newcommand{\pd}[1]{\mathcal{S}_{#1}^{+}(\R)}
\newcommand{\gln}[1]{\textnormal{GL}_{#1}}
\newcommand{\id}{\mathbb{I}}
\newcommand{\diag}[1]{\textnormal{diag}(#1)}

\newcommand{\avg}{\mathbb{E}}
\newcommand{\E}{\avg}

\newcommand{\dd}{\textrm{d}}
\newcommand{\bm}{}
\newcommand{\bp}{{\bm p}}
\newcommand{\bP}{{\bm P}}
\newcommand{\bq}{{\bm q}}
\newcommand{\bQ}{{\bm Q}}

\newcommand{\bmu}{{\bm \mu}}
\newcommand{\bmup}{\bmu_p}
\newcommand{\bmuP}{\bmu_P}

\newcommand{\tbmup}{\tilde{\bmu}_p}
\newcommand{\tbmuP}{\tilde{\bmu}_P}

\newcommand{\bw}{{\bm W}}

\newcommand{\bsigma}{\bm{\sigma}}

\newcommand{\cov}{\text{Cov}}

\newcommand{\vega}{\mathcal{V}}
\newcommand{\bdelta}{\bm{\Delta}}
\newcommand{\bdeltafull}{\bdelta_{\textnormal{c}}}
\newcommand{\bvegafull}{\bvega_{\textnormal{c}}}

\DeclareMathAlphabet\mathbfcal{OMS}{cmsy}{b}{n}
\newcommand{\bvega}{\mathcal{V}}
\newcommand{\bvarsigma}{{\bm \varsigma}}

\newcommand{\volmatrix}{\mathcal{G}}
\newcommand{\volmatrixp}{\mathcal{G}_p}
\newcommand{\volmatrixP}{\mathcal{G}_P}

\newcommand{\ofdrift}{\nu}

\newcommand{\ofdiff}{\mathcal{L}}
\newcommand{\ofdiffqq}{\ofdiff_{qq}}
\newcommand{\ofdiffqQ}{\ofdiff_{qQ}}
\newcommand{\ofdiffQq}{\ofdiff_{Qq}}
\newcommand{\ofdiffQQ}{\ofdiff_{QQ}}

\newcommand{\Omegaqq}{\Omega_{qq}}
\newcommand{\OmegaqQ}{\Omega_{qQ}}
\newcommand{\OmegaQq}{\Omega_{Qq}}
\newcommand{\OmegaQQ}{\Omega_{QQ}}
\newcommand{\OmegaXi}{\Omega_{\Xi}}

\newcommand{\Sigmapp}{\Sigma_{pp}}

\newcommand{\agg}{\textnormal{Agg}}

\newcommand{\cimpact}{\Lambda}
\newcommand{\cimpactpq}{\Lambda_{pq}}
\newcommand{\cimpactPq}{\Lambda_{Pq}}
\newcommand{\cimpactpQ}{\Lambda_{pQ}}
\newcommand{\cimpactPQ}{\Lambda_{PQ}}

\newcommand{\lambdadirect}{\Lambda_{\textnormal{direct-4d}}}
\newcommand{\lambdabs}{\Lambda_{\textnormal{bs}}}
\newcommand{\lambdatwod}{\Lambda_{\textnormal{2d}}}
\newcommand{\lambdatwodnox}{\Lambda_{\textnormal{direct-2d}}}
\newcommand{\lambdafourd}{\Lambda_{\textnormal{4d}}}

\newcommand{\rsqin}{\mathcal{R}^2_{\text{in}}}

\newcommand{\Lc}{\mathcal{L}}

\newcommand{\covar}[1]{\langle #1 \rangle}

\newtheorem{assumptionalt}{Assumption}[assumption]
\newtheorem{lemmaalt}{Lemma}[lemma]
\newtheorem{propositionalt}{Proposition}[proposition]
\crefname{assumptionalt}{Assumption}{Assumptions}
\crefname{propositionalt}{Proposition}{Propositions}
\crefname{lemmaalt}{Lemma}{Lemmas}

\newenvironment{lemmap}[1]{
  \renewcommand\thelemmaalt{#1}
  \lemmaalt
}{\lemmaalt}

\newenvironment{propositionp}[1]{
  \renewcommand\thepropositionalt{#1}
  \propositionalt
}{\endassumptionalt}

\newcommand{\prefactor}{Y}

\title{Cross impact in derivative markets \thanks{The authors warmly thank J.-P. Bouchaud, Z. Eisler and B. Toth and M. Rosenbaum for fruitful discussions. This research was conducted within the \textit{Econophysics \& Complex Systems} Research Chair under the aegis of the  \textit{Fondation du Risque}, a joint initiative by the \textit{Fondation de l'\'Ecole polytechnique, l'\'Ecole polytechnique} and Capital Fund Management. Mehdi Tomas gratefully acknowledges financial support of the ERC 679836 Staqamof and the Chair Analytics and Models for financial regulation, Deep finance and statistics, Machine learning and systematic methods.}}

\author[1,2,3]{\normalsize Mehdi Tomas}
\author[3,4]{Iacopo Mastromatteo}
\author[1,3,4]{Michael Benzaquen}
 
\affil[1]{\normalsize LadHyX UMR CNRS 7646, Ecole Polytechnique, 91128 Palaiseau Cedex, France}
\affil[2]{\normalsize CMAP UMR CNRS 7641, Ecole Polytechnique, 91128 Palaiseau Cedex, France}
\affil[3]{\normalsize Chair of Econophysics \& Complex Systems, Ecole Polytechnique, 91128 Palaiseau Cedex, France}
\affil[4]{\normalsize Capital Fund Management, 23-25, Rue de l’Universit\'e 75007 Paris, France}

\begin{document}
\maketitle

\begin{abstract}
    Trading a financial asset pushes its price as well as the prices of other assets, a phenomenon known as cross-impact. The empirical estimation of this effect on complex financial instruments, such as derivatives, is an open problem. To address this, we consider a setting in which the prices of \emph{derivatives} is a deterministic function of stochastic \emph{factors} where trades on both factors and derivatives induce price impact. We show that a specific cross-impact model satisfies key properties which make its estimation tractable in applications. Using E-Mini futures, European call and put options and VIX futures, we estimate cross-impact and show our simple framework successfully captures some of the empirical phenomenology. Our framework for estimating cross-impact on derivatives may be used in practice for estimating hedging costs or building liquidity metrics on derivative markets. 
\end{abstract}

\section*{Introduction}

\textit{Market impact} describes how trades on one asset translate into its price. The many studies on the topic have deepened our understanding of how markets digest trades into prices \cite{Bouchaud2018TradesPrices,Almgren2005DirectImpact,Torre1997BARRAHandbook}. However, market impact ignores the effects of order flows of other assets on an asset's price. This phenomenon, dubbed \textit{cross-impact}, has received growing attention in recent years \cite{Hasbrouck2001CommonLiquidity,Pasquariello2015StrategicMarket,Schneider2017Cross-impactNo-dynamic-arbitrage,Wang2015PriceResults,tomas2020build,rosenbaum2021characterisation}. Estimating models which account for cross-impact poses new challenges and led the authors to investigate in \cite{tomas2020build} which cross-impact models satisfy desirable properties while fitting empirical data well.
\\ \\
This paper contributes to the literature on estimating cross-impact by focusing on a particular asset class: derivatives. In this asset class, cross-sectional impact effects are expected to play a major role. Indeed, in an efficient market, the prices of derivatives should be locked by no-arbitrage. Therefore, the impact of trades on different derivatives should be related. Furthermore, since there are thousands of strongly correlated and individually illiquid derivatives, we need to aggregate the liquidity of instruments to properly estimate impact.
\\ \\
We set up a continuous-time, state-dependent static cross-impact framework which generalises the setup of \cite{tomas2020build}. In this framework, derivative prices are functions of stochastic factors and trades impact prices of all instruments. The main result of this paper is a calibration formula for cross-impact on derivatives, which we find provides better goodness-of-fit than simple impact models.
\\ \\
We now comment on the links between our approach and the literature.
\\ \\
The methodology of this paper is most similar to a previous paper on cross-impact by the authors \cite{tomas2020build}. This paper introduces an infinitesimal extension of the framework of the previous paper in which cross-impact also depends on state variables (such as factor prices). These are slight modifications to handle the specific challenges posed by derivatives.
\\ \\
The cross-impact model used in this paper is the Kyle cross-impact model, first derived using a multivariate version of the Kyle insider trading framework \cite{Caballe1994ImperfectNeutrality,kyle1985continuous,delMolino2018TheDifferent}. The properties of this model were examined in \cite{tomas2020build} and this paper leverages some results from \cite{tomas2020build}.
\\ \\
This paper focuses on linearly relating anonymous trade data to option prices. This linear approximation is roughly justified for anonymous order flow but not for a particular investor's trades. The papers \cite{Toth2016TheMarkets,Said2019HowImpact} study this question by using databases of proprietary investor orders to estimate the price impact of trades on the implied volatility surface. In particular, \cite{Toth2016TheMarkets} shows that option trades have a vega-weighted impact on the level of the implied volatility surface. The work \cite{Said2019HowImpact} finds that trades also impact the skew or term structure of the implied volatility surface. When applying our framework to empirical data, we recover a linear version of these two effects, along with an estimation of the importance of each effect's contributions.
\\ \\
Finally, this paper focuses on how market prices of options are affected by trades and ignores the intricacies of the replication problem. Thus, it is separate from works which model how price impact modifies option prices and the associated hedging strategies under different replication constraints \cite{Loeper2018OptionEquations,Bouchard2017HedgingConstraint}. However, the current framework may be useful to give empirical estimates to the impact coefficients appearing in these papers.
\\ \\
The paper is organized as follows. \cref{sec:setup} presents our modeling framework. \cref{sec:properties_impact} derives a tractable calibration formula for cross-impact. \cref{sec:examples} provides illustrative examples of the framework. \cref{sec:empirical} applies the setup to estimate cross-impact on options. Finally, we conclude on the contributions of the paper, open questions, and directions for future work.

\section*{Notation}
\label{sec:nota}

The set of $n$ by $n$ real-valued square matrices is denoted by $\mat{n}$, the set of orthogonal matrices by $\orth{n}$, the set of real non-singular $n$ by $n$ matrices by $\gln{n}(\reals{})$, the set of real $n$ by $n$ symmetric positive semi-definite matrices by $\spd{n}$, and the set of real $n$ by $n$ symmetric positive definite matrices by $\pd{n}$. Further, given a matrix $A$ in $\mat{n}$, $A^\top$ denotes its transpose. Given $A$ in $\spd{n}$, we write $A^{1/2}$ for a matrix such that $A^{1/2} (A^{1/2})^\top = A$ and $\sqrt{A}$ for the matrix square root, the unique positive semi-definite symmetric matrix such that $(\sqrt{A})^2 = A$.
\\ \\
All stochastic processes in the text are defined on a probability space $(\Omega, \mathcal{F}, (\mathcal{F}_t)_{t \in \reals{}}, \mathbb{P})$ and adapted to the filtration $(\mathcal{F}_t)_{t \in \reals{}}$ unless stated otherwise. Standard Brownian motions are defined with respect to the probability measure $\mathbb{P}$. All stochastic differential equations introduced will be assumed to have a unique strong solution and correspondingly the functions appearing in these equations will be assumed to be sufficiently regular for this to be true. We denote by $\avg$ the expectation with respect to the probability measure $\mathbb{P}$. We denote by $\covar{X,Y}$ the quadratic covariation of the continuous stochastic processes $X$ and $Y$.

\section{Setup}
\label{sec:setup}

We consider a universe comprising two classes of financial instruments, that we will refer to as \emph{factors} and \emph{derivatives}. 
\\ \\
Factors represent a set of $N$ stochastic processes from which one can derive derivative prices. These may include the underlying of derivative contracts, such as the spot, as well as stochastic or local volatility, forward variance, yield curves, etc. The prices of factors at time $t$ is denoted by $\bp_t =(p^1_t, \dots, p^N_t)$. Derivatives are a set of $M$ contracts, whose prices at maturity depend on the values of the factors. We write $\bP_t = (\bP^1_t, \dots, \bP^M_t)$ for the prices of these instruments.
\\ \\
Factors and derivatives are traded continuously and we denote by $\bq_t = (q^1_t, \dots, q^N_t)$ the stochastic process corresponding to the market net traded order flows on factors and by $\bQ_t = (Q^1_t, \dots, Q^M_t)$ the stochastic process corresponding to the market net traded order flows on derivatives. These are aggregate order flow, corresponding to the sum of all trades sent by market participants.
\\ \\
As we are interested in a simple, infinitesimal framework for the influence of trades on prices, we assume that order flow dynamics on each asset are continuous stochastic processes driven by Brownian motions.
\begin{assumption}[Order flow dynamics]
\label{ass:of_dynamics}
The order flow follows the following stochastic dynamics
\begin{align}
    \dd \bq_t &= \ofdrift_q(\bp_t, t) \dd t + \ofdiff_{qq}(\bp_t, t) \dd Z^q_t + \ofdiff_{Qq}(\bp_t, t) \dd Z^Q_t \\
    \dd \bQ_t &= \ofdrift_Q(\bp_t, t) \dd t + \ofdiff_{Qq}(\bp_t, t) \dd Z^q_t + \ofdiff_{QQ}(\bp_t, t) \dd Z^Q_t \, ,
\end{align}
where $Z^q, Z^Q$ are uncorrelated standard Brownian motions, $\ofdrift_q \colon \reals{N} \times \reals{} \to \reals{N}$ and $\ofdrift_Q \colon \reals{N} \times \reals{} \to \reals{M}$ encode order flow drift, $\ofdiff_{qq} \colon \reals{N} \times \reals{} \to \matsq{N}$, $\ofdiff_{QQ} \colon \reals{N} \times \reals{} \to \matsq{M}$, $\ofdiff_{Qq} \colon \reals{N} \times \reals{} \to \mat{M,N}$, $\ofdiff_{qQ} \colon \reals{N} \times \reals{} \to \mat{N,M}$ encode co-trading among assets.
\end{assumption}

\cref{ass:of_dynamics} allows for correlations of order flows between assets, so that our model accounts for co-trading of factor and derivatives, which is a typical feature expected in derivative markets. However, the continuous modeling framework for order flows \cref{ass:of_dynamics} is somewhat unrealistic. Indeed, at the high-frequency scale, trades arrive punctually and trade arrivals clustered so that a more realistic modeling is driven by point processes which respect this property. Such modeling requires lengthy mathematical treatment and is outside the scope of this paper, so that we choose a simple continuous order flow model. We refer the reader to \cite{rosenbaum2021characterisation} for a model which accounts for these effects.
\\ \\
We now move to the dynamics of factor prices in our setup.

\begin{assumption}[Factor dynamics]
\label{ass:prices}
We assume that factor prices are given by
\begin{align}
\label{eq:underlying_dynamics}
\dd \bp_t &= \bmup(\bp_t, t) \dd t + \volmatrixp(\bp_t, t) \dd \bw_t + \cimpactpq(\bp_t, t) \dd \bq_t + \cimpactpQ(\bp_t, t) \dd \bQ_t
\end{align}
where $\bw$ is a standard $N$-dimensional Brownian motion, $\bmup  \colon \reals{N} \times \reals{} \to \reals{N}$ is the factor drift, $\volmatrix_p \colon \reals{N} \times \reals{} \to \gln{N}(\reals{})$ is the factor diffusion matrix, and $\Lambda_{pq} \colon \reals{N} \times \reals{} \to \mat{N}$, $\Lambda_{pQ} \colon \reals{N} \times \reals{} \to \mat{N,M}$ capture cross-impact.
\end{assumption}

The factor dynamics of \cref{ass:prices} are quite general. They encompass local and stochastic volatility models and allow for cross-impact between factors and between factors and derivatives. We assume that factors are chosen so that derivatives are priced in a Markovian manner with respect to the factors, which leads us to the next assumption.

\begin{assumption}[Derivative prices]
\label{ass:derivative_function}
There exists a function $F \colon \R^N \times \R \to \R^M$, twice differentiable with respect to the first argument and differentiable with respect to the second argument, such that $\bP_t = F(\bp_t, t)$.
\end{assumption}

Applying Ito's formula to \cref{eq:underlying_dynamics} and using \cref{ass:derivative_function,ass:of_dynamics}, we obtain the following corollary.

\begin{corollary}
\label{lemma:derivative_ito}
The derivative dynamics are given by
\begin{align}
\label{eq:derivative_dynamics}
\dd \bP_t &= \bmuP(\bp_t, t) \dd t + \volmatrixP(\bp_t, t) \dd \bw_t + \cimpactPq(\bp_t, t) \dd \bq_t + \cimpactPQ(\bp_t, t) \dd \bQ_t \, ,
\end{align}
where $\bmuP  \colon \reals{N} \times \reals{} \to \reals{M}$ is the derivative drift, $\volmatrix_P \colon \reals{N} \times \reals{} \to \mat{M,N}$ is the derivative diffusion matrix, $\Lambda_{Pq} \colon \reals{N} \times \reals{} \to \mat{M,N}$ and $\Lambda_{PQ} \colon \reals{N} \times \reals{} \to \mat{M}$ encode cross-impact. In particular, we have the constraints $\cimpactPq = \Xi \cimpactpq$, $\cimpactPQ = \Xi \cimpactpQ$, $\volmatrixP = \Xi \volmatrixp$, where $\Xi := (\frac{\partial \bP^i}{\partial \bp^j})_{i,j}$ is the $M$ by $N$ sensitivity matrix.
\end{corollary}

\cref{lemma:derivative_ito} does not make explicit the dependence of the derivative drift on other variables as it will not play an important role. It further shows that the matrices $\cimpactPq$ and $\cimpactPQ$ are constrained, but not $\cimpactpq$ or $\cimpactpQ$.
\\ \\
For convenience cross-impact matrices appearing in \cref{eq:underlying_dynamics,eq:derivative_dynamics} can be compactly rearranged into a single matrix, $\Lambda$, which we refer to as the cross-impact matrix since it describes the cross-impact of the complete system
\begin{equation}
    \label{eq:grouped_impact}
    \Lambda(\bp_t, t) :=  \begin{pmatrix} \Lambda_{pq} & \Lambda_{pQ} \\ \Lambda_{Pq} & \Lambda_{PQ} \end{pmatrix} (\bp_t, t) \, .
\end{equation}
The impact model we propose involves two parameters, the return covariance matrix and the order flow covariance matrix. 
\\ \\
The factor-factor return covariance matrix $\Sigma_{pp} \colon \reals{N} \times \reals{} \to \spd{N}$ is defined as
\begin{equation}
\label{eq:covariance_udr}
\Sigma_{pp}(\bp_t, t) \dd t := 
\dd \covar{\bp, \bp}_t \, ,
\end{equation}
and we similarly denote $\Sigma_{pP}, \Sigma_{Pp}=\Sigma_{pP}^\top, \Sigma_{PP}$ for the factor-derivative, derivative-factor and derivative-derivative return covariance matrices. Naturally, since derivative prices are deterministic function of factors, these matrices are all related to $\Sigma_{pp}$. 
\\ \\
We denote by $\Omega_{qq} \colon \reals{N} \times \reals{} \to \spd{N}$ the factor-factor order flow covariance matrix
\begin{equation}
\label{eq:covariance_qq}
\Omega_{qq}(\bp_t, t) \dd t := 
\dd \covar{\bq, \bq}_t \, ,
\end{equation}
and we denote $\Omega_{qQ}$, $\Omega_{Qq}=\Omega_{qQ}^\top$ and $\Omega_{QQ}$ for the factor-derivative, derivative-factor and derivative-derivative order flow covariances. Contrary to return covariance matrices, there are no constraints betweeen these order flow covariance matrices and $\Omega_{qq}$. The covariance structure of returns and flows for the whole system can be arranged compactly as
\begin{equation*}
  \Sigma(\bp_t, t) = \begin{pmatrix} \Sigma_{pp} & \Sigma_{pP} \\  \Sigma_{pP}^\top & \Sigma_{PP}  \end{pmatrix} (\bp_t, t)
  \hspace{1cm}
  \Omega(\bp_t, t) = \begin{pmatrix} \Omega_{qq} & \Omega_{qQ} \\ \Omega_{qQ}^\top & \Omega_{QQ} \end{pmatrix} (\bp_t, t) \, .
\end{equation*}

The cross-impact matrix that we propose to use, dubbed the \emph{Kyle cross-impact model}, was first derived in \cite{Caballe1994ImperfectNeutrality} and then analyzed in \cite{delMolino2018TheDifferent,tomas2020build}. The next assumption makes explicit the formula of the Kyle cross-impact matrix within our framework.

\begin{assumption}[Cross-impact matrix]
\label{ass:impact_model}
The cross-impact matrix $\cimpact$ is of the form
\begin{equation}
    \label{eq:kyle_model}
    \cimpact := \sqrt{\prefactor} (\Lc^{-1})^\top \sqrt{\Lc^\top \Sigma \Lc} \Lc^{-1} \, ,
\end{equation}
where $\Lc$ is a matrix such that $\Lc \Lc^\top = \Omega$, $\prefactor$ is a constant such that $0 < \prefactor < 1$, and we have omitted the dependence on $(\bp_t, t)$ for compactness.
\end{assumption}

The choice of the Kyle cross-impact matrix is motivated by the goodness-of-fit observed in previous work \cite{tomas2020build} and some key properties it satisfies, which will be outlined in the next section. Note that the cross-impact matrix of \cref{ass:impact_model} must, at the very least, be compatible with the constraints on the cross-impact matrix outlined in \cref{lemma:derivative_ito}. We will show that this is true in the next section.
\\ \\
Our framework being set up, we move to the derivation of key properties of our cross-impact framework in the next section. These properties motivate the choice of the Kyle cross-impact model as the only well-behaved model for cross-impact on derivatives.

\section{A practical formula for the cross-impact matrix}
\label{sec:properties_impact}

This section shows how one can derive the cross-impact matrix $\cimpact$ in a tractable form for applications.
\\ \\
The next proposition shows that the impact contributions to factor and derivative prices can be absorbed inside Brownian motions. 

\begin{proposition}
\label{prop:factoring_of}
Using the notation of \cref{ass:impact_model,lemma:derivative_ito,ass:prices}, we have
\begin{align}
\dd \bp_t &= \tbmup(\bp_t, t) \dd t + \frac{1}{\sqrt{1-\prefactor}} \volmatrixp(\bp_t, t) \dd B_t \\
\dd \bP_t &= \tbmuP(\bp_t, t) \dd t + \frac{1}{\sqrt{1-\prefactor}} \volmatrixP(\bp_t, t) \dd B_t \, ,
\end{align}
where $\tbmup  \colon \reals{N} \times \reals{} \to \reals{N}$, $\tbmuP  \colon \reals{N} \times \reals{} \to \reals{M}$ and $B$ is a standard $N$ dimensional Brownian motion.
\end{proposition}

The proof of \cref{prop:factoring_of} is given in \cref{app:proofs}. \cref{prop:factoring_of} underscores a convenient property of our framework: the sensitivities of derivative prices with respect to factors are independent of the order flow dynamics. In particular, the Greeks can be computed using traditional derivative pricing methods since derivative prices satisfy the classic stochastic differential system of \cref{prop:factoring_of}.
\\ \\
The next proposition shows that the large cross-impact matrix $\cimpact$ can be expressed solely as a function of the derivative sensitivities and the compact cross-impact matrix $\cimpactpq$.

\begin{proposition}
\label{prop:arbitrage}
We have
\begin{equation}
\label{eq:kyle_fragm}
\Lambda(\bp_t, t) = \begin{pmatrix} \Lambda_{pq} & \Lambda_{pq} \Xi^\top \\ \Xi \Lambda_{pq} & \Xi \Lambda_{pq} \Xi^\top \end{pmatrix} (\bp_t, t) \, ,
\end{equation}
where we recall that $\Xi := (\frac{\partial \bP^i}{\partial \bp^j})_{i,j}$ is the $M$ by $N$ sensitivity matrix.
\end{proposition}

The proof of \cref{prop:arbitrage} is given in \cref{app:proofs}. Note that \cref{prop:arbitrage} shows that the chosen cross-impact model satisfies the constraints of \cref{lemma:derivative_ito}. Furthermore, it expresses the large $N+M$ by $N+M$ cross-impact matrix $\cimpact$ as a function of the much smaller $N$ by $N$ cross-impact matrix $\cimpactpq$ and the derivative sensitivities $\Xi$. Thanks to \cref{prop:factoring_of}, the latter can be computed as the usual Greeks of our derivative pricing model. Thus, given a formula for $\cimpactpq$, the cross-impact matrix $\cimpact$ can be computed. The next proposition proves that the cross-impact matrix $\cimpactpq$ can be expressed as a function of the factor return covariance matrix $\Sigma_{pp}$ and a greek-weighted covariance matrix of order flows, denoted by $\OmegaXi$. The latter aggregates factor and derivative liquidity in a smaller $N$ by $N$ matrix. 

\begin{proposition}
\label{prop:kyle_short}
We have
\begin{equation}
\label{eq:kyle_short}
\Lambda_{pq} = \sqrt{\prefactor} (\Lc_{\Xi}^{-1})^\top \sqrt{\Lc_{\Xi}^\top \Sigma_{pp} \Lc_{\Xi}} \Lc_{\Xi}^{-1} \, ,
\end{equation}
where we have omitted the dependence on $(\bp_t, t)$, $\OmegaXi := \Omegaqq + \Xi^\top \OmegaQQ \Xi + \Xi^\top \OmegaQq + \OmegaqQ \Xi$, and $\Lc_{\Xi}$ is a matrix such that $\Lc_\Xi \Lc_\Xi^\top = \OmegaXi$.
\end{proposition}

The proof of \cref{prop:kyle_short} is given in \cref{app:proofs}. Combined, \cref{prop:kyle_short,prop:arbitrage} give a formula for the cross-impact matrix $\cimpact$ as a function of the measurable $N$ by $N$ matrices $\Sigma_{pp}$ and $\Omega_{\Xi}$. Furthermore, by \cref{prop:factoring_of}, $\Xi$ can be computed using usual derivative pricing methods. Overall, we have thus derived a scheme for estimating a cross-impact matrix on derivatives.
\\ \\
A relevant insight of \cref{eq:kyle_short} for applications is that even if factors are not traded, as long as derivatives with sensitivities to these factors are traded, i.e. $\Xi^\top \OmegaQQ \Xi$ is positive definite, then the inverses appearing in \cref{eq:kyle_short} are well-defined. This is not obvious from the form of the Kyle cross-impact matrix in \cref{ass:impact_model}. This property is important for applications where most factors correspond to non-tradeable instruments, such as volatility factors.

\section{Examples}
\label{sec:examples}

To illustrate the flexibility of our setup and the usefulness in practice of the properties of the Kyle cross-impact model, we discuss examples in increasing complexity below.

\subsection{Futures}

For our first example, we consider a universe of $N=M=1$ instruments, consisting in a spot with price $p_t$ and a futures contract delivering one unit of the spot and expiring at a later time $T$. By assuming a constant, continuously compounded, deterministic interest rate $r$ the derivative price is given by
$$
P(p_t, t) = e^{r(T-t)} p_t \, .
$$
Therefore, in this case $\Xi(p_t, t) = \partial_p P(p_t, t) = e^{r(T-t)}$ and \cref{eq:kyle_fragm} yields
$$
\Lambda(p_t, t) = \sqrt{\prefactor} p_t \frac{\sigma(p_t, t)}{\omega(p_t, t)} \begin{pmatrix} 1 & \ e^{r(T-t)} \\ e^{r(T-t)} & e^{2r(T-t)} \end{pmatrix},
$$
where $\sigma^2(p_t, t) \dd t := \dd \covar{p,p}_t$ and $\omega^2(p_t, t) := (1, e^{r(T-t)})^\top \Omega(p_t, t) (1, e^{r(T-t)})$. The meaning of this formula is simple: in a constant interest-rate model, the market liquidity should mix liquidity traded the spot and on the future, after properly adjusting for the interest rate.

\subsection{Black-Scholes model}
\label{sec:black_scholes_model}
We now consider a system with a single factor ($N=1$) and $M$ derivatives. The factor is the spot with price $p_t$ and the derivatives are a set of European call or put options with different strikes and maturities. We assume that the spot price $p_t$ follows log-normal dynamics with cross-impact, given by
$$
\dd p_t = \mu p_t \dd t + \sqrt{1-\prefactor} \sigma  p_t \dd \bw_t + \cimpactpq \dd \bq_t + \cimpactpQ \dd \bQ_t \, ,
$$
where $\bw_t$ is a one-dimensional standard Brownian motion, $\mu$ is the drift and $\sigma$ is the implied volatility. The parameter $\sigma$ is the implied volatility since \cref{prop:factoring_of} implies that there exists some drift $\tilde \mu$ such that
$$
\dd p_t = \tilde \mu p_t \dd t + \sigma p_t \dd B_t \, ,
$$
where $B$ is a one-dimensional standard Brownian motion. Then, with the usual notation for the Black-Scholes $\Delta$, we have
$$
\Xi^i(p_t, t) = \partial_p P^i(p_t, t) = \Delta^i(p_t, t) \, ,
$$
and, writing $\bdelta := (\partial_p P^1(p_t, t), \cdots, \partial_p P^M(p_t, t))$, \cref{eq:kyle_fragm} yields
$$
\Lambda(p_t, t) = \sqrt{\prefactor} p_t \frac{\sigma}{\omega(p_t, t)} \begin{pmatrix} 1 & \ \bdelta^\top \\ \bdelta & \bdelta \bdelta^\top  \end{pmatrix} (p_t, t) \, ,
$$
where
$$
\omega^2(p_t, t) = (1, \bdelta(p_t, t))^\top \Omega(p_t, t) (1, \bdelta(p_t, t)) \, .
$$
Thus, as in the previous example, there is a single liquidity pool, with volumes traded on options adjusted for the options' $\Delta$. Volume traded on deep in-the-money options ($\Delta^i \approx 1$) contribute to the overall liquidity pool as if it was the spot itself that was traded, whereas deeply out-of-the-money options ($\Delta^i \approx 0$) give negligible contributions.

\subsection{Volatility factors}
\label{sec:vol_factors_example}

\subsubsection{General setup}

We build on our previous example and consider a spot and a strip of $M$ European call and put options, with different implied volatilities for each option. The spot price in this example is denoted by $s_t$, while option prices are given by $P^i_t(s_t, \hat \sigma^i_t)$ where $\hat \sigma^i_t$ is the implied volatility of this option.
\\ \\
In order to reduce the dimensionality of the $M$ implied volatilities $\hat \sigma^i_t$, we assume that they are completely described by a set of volatility factors $\bvarsigma_t = (\varsigma^1_t, \dots,  \varsigma^Q_t)$ such that the implied volatility of each option is given by
\begin{equation}
\label{eq:volatility_factor_implied}
\hat \sigma^i_t = F^i(\bvarsigma_t) \, ,
\end{equation}
where $F^i \colon \R^Q \to \R$ is some function of these factors. With some abuse of notation, we write
$$
P^i(s_t, \hat F^i(\bvarsigma_t), t) =
P^i(s_t, \bvarsigma_t, t) \, ,
$$
and we will employ a similar notation for other functions of the implied volatility $\hat \sigma^i_t$. Our instruments thus comprise $N = 1 + Q$ factors, of which only one is tradeable (the spot), and where the other $Q$ factors correspond to non-tradeable volatility factors. The sensitivities of the system in this case correspond to
\begin{align*}
\Xi^{i1}_t (s_t, \bvarsigma_t, t) 
&=
\Delta^i_t (s_t, \bvarsigma_t, t)
:=
\frac{
    \partial P^i(s_t, \bvarsigma_t, t)
    }{\partial s} \\
\Xi^{i(q+1)}_t (s_t, \hat \sigma^i_t, t)
&=
\frac{
    \partial P^i(s_t, \hat\sigma^i_t, t)
    }{\partial \hat\sigma^i}
\frac{
    \partial F^i(\bvarsigma_t)
    }{\partial \varsigma^q} 
    :=
    \vega^i_t (s_t, \bvarsigma_t, t)
    \beta^{iq}(s_t, \varsigma_t, t) =: \Upsilon^{iq}
    \, ,
\end{align*}
where $q=1, \dots, Q$ and where, as it is customary in the literature on option pricing, we have introduced the \emph{vega}
$$
\vega^i_t (s_t, \hat \sigma^i_t, t) = 
\frac{
    \partial P^i_t(s_t, \hat\sigma^i_t, t)
    }{\partial \hat\sigma^i} \, ,
$$
and the sensitivities of the volatility surface to $ \bvarsigma_t$
$$
\beta^{iq}(s_t, \bvarsigma_t, t) = \frac{
    \partial F^i(\bvarsigma_t)
    }{\partial \varsigma^q} \, .
$$

This simple setup allows us to capture some of the salient implied volatility surface dynamics and we will make use of it in the next section.

\subsubsection{Single factor model}

This section examines the particular case when the volatility surface depends on a single volatility factor, i.e. $Q=1$. The following lemma shows that given some additional assumptions, we can derive an explicit formula for $\cimpactpq$.

\begin{lemma}
\label{lemma:kyle_leverage}
We denote by $\bdeltafull := (1, 0, \frac{\partial P^1}{\partial s}, \cdots, \frac{\partial P^M}{\partial s})$, $\bvegafull := (0, 1, \frac{\partial P^1}{\partial \varsigma}, \cdots, \frac{\partial P^M}{\partial \varsigma})$ the vectors of sensitivities. Then, if $\bdeltafull^\top \Omega \bvegafull = 0$, we have
\begin{equation}
\label{eq:kyle_leverage}
\cimpactpq(s_t, \varsigma_t, t) = \dfrac{\sqrt{\prefactor}}{\sqrt{\sigma^2 \omega_{\Delta}^2 + \xi^2 \omega_{\vega}^2 + 2 \sigma \xi \rho \omega_{\Delta} \omega_{\vega}}}
\begin{pmatrix}
\sigma^2 + \frac{\omega_{\vega}}{\omega_{\Delta}} \sigma \xi \sqrt{1-\rho^2} & \sigma \xi \rho \\
\sigma \xi \rho & \xi^2 + \frac{\omega_{\Delta}}{\omega_{\vega}} \sigma \xi \sqrt{1-\rho^2}
\end{pmatrix} (s_t, \varsigma_t, t) \, ,
\end{equation}
where $\omega_{\Delta}^2 := \bdeltafull^\top \Omega \bdeltafull$ is the delta-aggregated liquidity, $\omega_{\vega}^2 := \bvegafull^\top \Omega \bvegafull$ is the vega-aggregated liquidity, $\sigma^2(s_t, \varsigma_t, t) \dd t := \dd \covar{s,s}_t$ is the spot volatility, $\xi^2(s_t, \varsigma_t, t) \dd t:= \dd \covar{\varsigma, \varsigma}_t$ is the volatility of volatility and $\rho(s_t, \varsigma_t, t) \dd t := \dfrac{\dd \covar{\varsigma, s}_t}{\xi(s_t, \varsigma_t, t) \sigma(s_t, \varsigma_t, t)}$ is the spot-vol correlation.
\end{lemma}

The proof of \cref{lemma:kyle_leverage} is given in \cref{app:proofs}. An interesting result of \cref{eq:kyle_leverage} is that a delta-hedged trade induces impact on the spot because of the negative spot-vol correlation. We will make use of this single factor model in the next section.

\section{Empirical Results}
\label{sec:empirical}

We now illustrate our setup with an empirical analysis of cross-impact on derivatives markets which makes use of the results derived in \cref{sec:properties_impact}. \cref{sec:setupemp} describes the universe of instruments and the chosen derivative modeling. \cref{sec:relevant_observables} shows the empirical observables $\Sigma_{pp}$ and $\Omega_{\Xi}$ used in \cref{sec:cross-impact-models} to compute the resulting cross-impact matrix $\Lambda_{pp}$. Finally, \cref{sec:quality_fit} stress-tests the fit of cross-impact models and \cref{sec:aggregate_impact} examines non-parametric evidence of cross-impact.

\subsection{Setup}
\label{sec:setupemp}

\begin{figure}[h!]
	\centering
		\centering
		\includegraphics[width=0.95\textwidth]{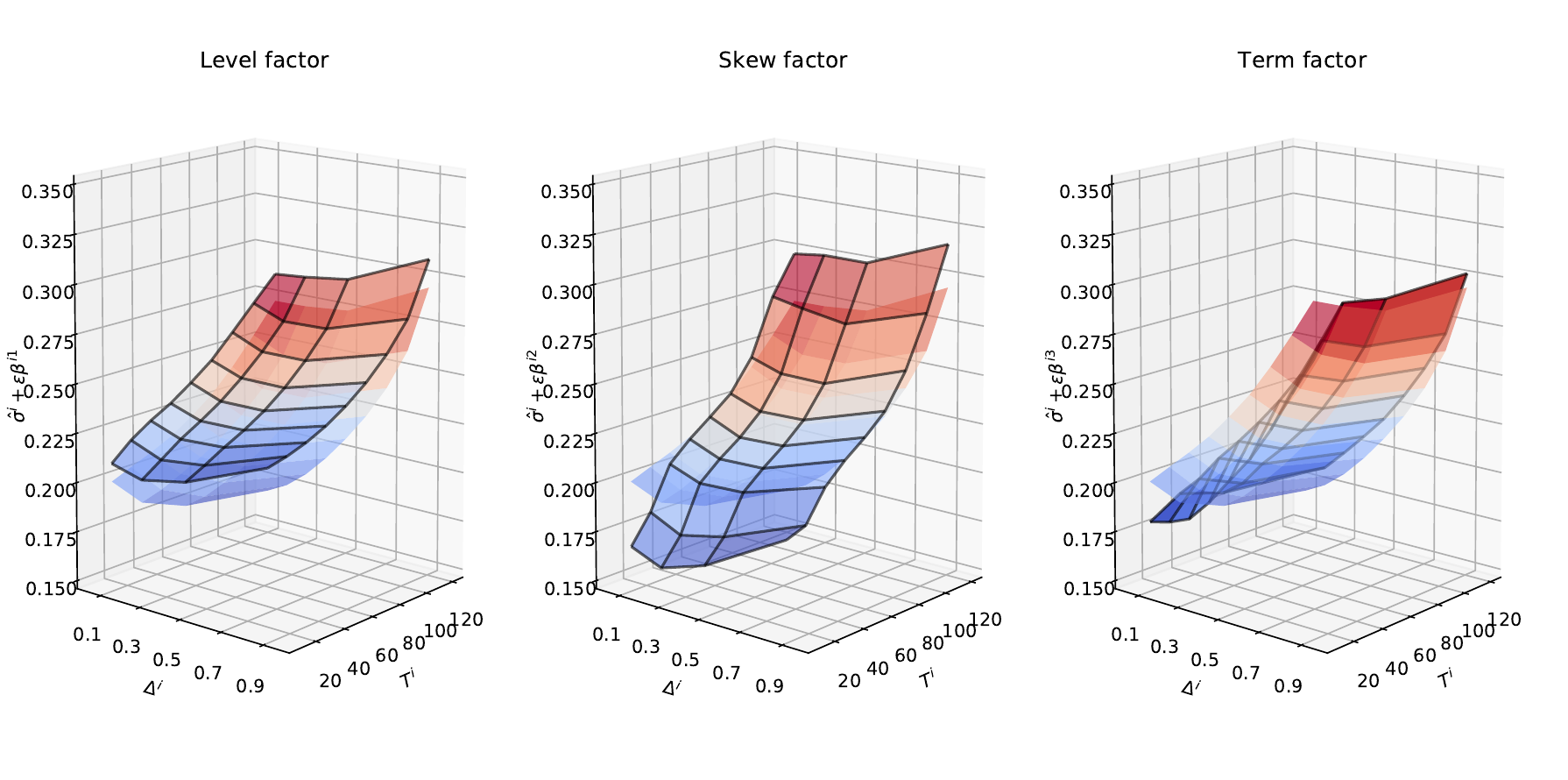}
		\caption{\textbf{Effect of the different volatility factors on the implied volatility surface.} \\
		Starting from a historical implied volatility surface $(\hat{\sigma}^i)$, we show the modified implied volatility surface $(\hat{\sigma}^i + \varepsilon \beta^{iq})$ after adding a contribution of size $\varepsilon$ from the volatility factor $q$. The original (non-modified) implied volatility surface is shown in light opacity for reference.}
	    \label{fig:volatility_factors}
\end{figure}

The universe of instruments is made up of (i) the front-month E-mini future, (ii) the two front-month VIX futures, (iii) a set of $M-2$ call and put options on the E-mini. We thus have $M$ derivatives. 
\\ \\
We bin returns and order flows on a time window $\Delta t$ of five minutes. We write $\delta \bp_t$ for the factor price change between time $t$ and time $t+\Delta t$, $\delta \bq_t$ for the signed order flow traded on factors within that time window and $\delta \bQ_t$ for the signed order flow traded on derivatives. Prices and order flows for these instruments are taken from trades and quotes data and more detail on the dataset is provided in \cref{app:data}.
\\ \\
We consider a linear approximation of the implied volatility surface with volatility factors, so that using the notations of \cref{sec:vol_factors_example}, we have $F^i(\bvarsigma) = \sum_{q=1}^{Q} \beta^{iq}\varsigma^q$ where $i=1,\dots,M$. To fit surfaces, we choose $Q=3$ and perform a principal component analysis of implied volatility surface returns (see, for example, \cite{Cont2002StochasticSurfaces}). These factors are given \cref{fig:volatility_factors}. The first volatility factor is a classic implied volatility level factor and we make the rough approximation that VIX futures are solely explained by such \emph{level} factor. The second volatility factor corresponds to the skew of the implied volatility surface, referred to as the \emph{skew} factor hereafter. The third volatility factor explains the term structure of the implied volatility, hence the name \emph{term} factor in the following.

\subsection{Estimated covariances}
\label{sec:relevant_observables}

\begin{figure}[h!]
	\centering
		\centering
		\includegraphics[width=0.5\linewidth]{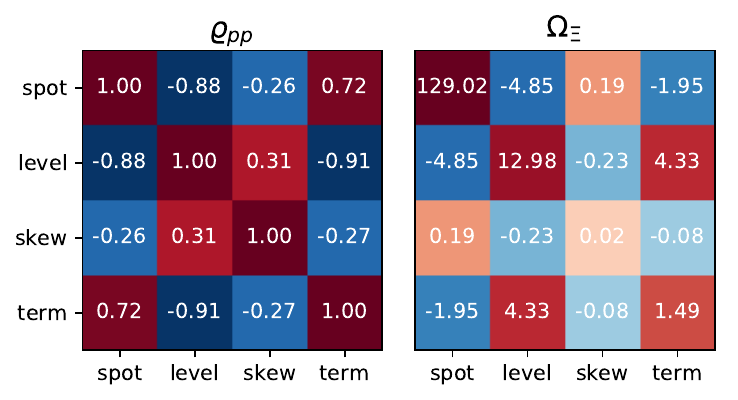}
		\caption{\textbf{Empirical estimates of the factor return correlation matrix $\varrho$ and aggregate order flow covariance $\OmegaXi$.} \\
		The factor return correlation matrix $\varrho$ (left) and the aggregate order flow covariance matrix $\Omega$ (right) estimates on our dataset. The order flow is reported in thousands of dollars of risk.}
	    \label{fig:covariance}
\end{figure}

\cref{prop:kyle_short} shows that we need to estimate $\Sigmapp$ and $\OmegaXi$ to compute $\cimpactpq$. We detail the estimation procedure in \cref{app:covariances}. \cref{fig:covariance} shows the estimated factor return correlation matrix $\varrho_{pp} := \diag{\bsigma}^{-1} \Sigma_{pp} \diag{\bsigma}^{-1}$ and the risk order flow covariance matrix $\Omega_{\Xi}^{\text{risk}}:= \diag{\bsigma} \Omega_{\Xi} \diag{\bsigma}$ where $\bsigma = ((\Sigma^{11}_{pp})^{1/2}, \cdots, (\Sigma_{pp}^{NN})^{1/2})$ is the factor volatility. 
\\ \\
The factor return correlation matrix correlation matrix shows strong negative correlation between the spot and level mode. This is a well-known stylised fact, sometimes referred to as the "leverage effect". This will play an important role in the form of the cross-impact model, as highlighted in \cref{eq:kyle_leverage_approx_emp}. Unsurprisingly, the correlation between spot and level order flow is much smaller, although still noticeable (around -0.15\%).
\\ \\
The traded risk (volatility times liquidity) is concentrated on the spot and level directions. This justifies approximating cross-impact on options using solely the spot and the level factor, which we delve in more detail in \cref{sec:cross-impact-models}. The traded risk in the skew direction is much smaller than all other directions and is thus expected to play a lesser role.

\subsection{Cross-impact models}
\label{sec:cross-impact-models}

\begin{figure}[h!]
	\centering
		\centering
		\includegraphics[width=0.3\linewidth]{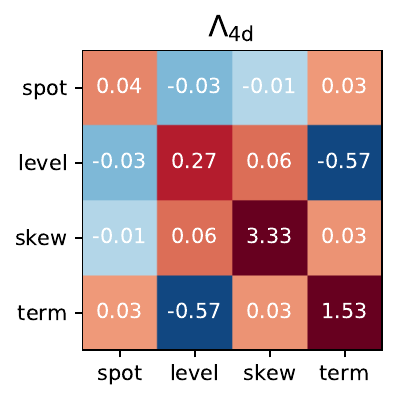}
		\caption{\textbf{Four dimensional Kyle cross-impact model on options.} \\
		We report the four dimensional Kyle model estimated using empirical estimates of the covariances of \cref{fig:covariance}. The cross-impact matrix are reported in units of risk and in basis points so that $\Lambda_{ij}$ encodes by how many basis points of volatility Asset $i$ is pushed by trading one dollar of risk on Asset $j$.}
	    \label{fig:kyle-4d}
\end{figure}

We can now use the empirical estimates of $\Sigma_{pp}$ and $\Omega_{\Xi}$ from the previous section to compute the cross-impact matrix $\Lambda$. For comparison purposes, we also introduce other choices of cross-impact matrices. All cross-impact matrices involve a prefactor $\prefactor$ which is calibrated to maximise goodness-of-fit. The first cross-impact matrix used for comparison is the Black-Scholes cross-impact model introduced in \cref{sec:black_scholes_model} which has a single factor: the spot. It is defined as
\begin{equation}
\lambdabs(\bp_t, t) := \sqrt{\prefactor} \frac{\sigma}{\sqrt{\bdeltafull^\top \Omega \bdeltafull}} \bdeltafull \bdeltafull^\top \, ,
\end{equation}
where $\bdeltafull := (1, 0, 0, 0, 0, 0, \Delta^3(\bp_t, t), \cdots, \Delta^{M}(\bp_t, t))$ is the delta vector, which places one on the spot, zero on the three volatility factors, zero on the two VIX futures, and the usual Black-Scholes delta on put and call options. The Black-Scholes model coincides with the Kyle cross-impact model if all the liquidity is concentrated on the spot. In particular, this model is unable to account for changes in the volatility factors. We thus introduce the two-dimensional direct model $\lambdatwodnox$ which accounts for the spot and implied volatility factor but ignores cross-sectional effects, defined as
\begin{equation}
\lambdatwodnox(\bp_t, t) := \sqrt{\prefactor} \frac{\sigma}{\sqrt{\bdeltafull^\top \Omega \bdeltafull}} \bdeltafull \bdeltafull^\top + \sqrt{\prefactor} \frac{\xi}{\sqrt{\bvegafull^\top \Omega \bvegafull}} \bvegafull \bvegafull^\top \, ,
\end{equation}
where $\bvegafull := (0, 1, 0, 0, 1, 1, \vega^3(\bp_t, t), \cdots, \vega^{M}(\bp_t, t)$ is the vega vector, which places zero on the spot, one on the level factor, zero on the other two volatility factors, and the usual Black-Scholes vega on put and call options. To account for all factor without correcting for cross-sectional effects, we introduce the four-dimensional direct model
\begin{equation}
\lambdadirect(\bp_t, t) := \sqrt{\prefactor} \frac{\sigma}{\sqrt{\bdeltafull^\top \Omega \bdeltafull}} \bdeltafull \bdeltafull^\top + \sqrt{\prefactor} \frac{\xi}{\sqrt{\bvegafull^\top \Omega \bvegafull}} \bvegafull \bvegafull^\top + \sqrt{\prefactor} \sum_{i=3}^{Q+1} \sqrt{\frac{\Sigma^{ii}_{pp}}{\Omega_{\Xi}^{ii}}} \Xi^{\cdot i}(\bp_t, t) (\Xi^{\cdot i}(\bp_t, t))^\top \, .
\end{equation}
Direct models ignore the off-diagonal structure of $\Sigma_{pp}$ and $\Omega_{\Xi}$. In particular they do not account for the leverage effect, which is an essential characteristic of the factor return covariance matrix $\Sigma_{pp}$. To fix this, we introduce the two-dimensional Kyle cross-impact model $\lambdatwod$ which captures the two dominating factor of the system: the spot and level factor. Since \cref{fig:covariance} shows that the delta and vega order flow correlation is small (around $-0.15\%$) and $\xi \omega_{\vega} \ll \sigma \omega_{\Delta}$, we can use \cref{eq:kyle_leverage} to obtain the approximation
\begin{equation}
\label{eq:kyle_leverage_approx_emp}
\lambdatwod(p_t, \varsigma_t, t) \approx \sqrt{\prefactor} \frac{\sigma}{\sqrt{\bdeltafull^\top \Omega \bdeltafull}} \bdeltafull \bdeltafull^\top  + \sqrt{\prefactor} \frac{\xi \sqrt{1-\rho^2}}{\sqrt{\bvegafull^\top \Omega \bvegafull}} \bvegafull \bvegafull^\top + \sqrt{\prefactor} \frac{\xi \rho}{\sqrt{\bdeltafull^\top \Omega \bdeltafull}} (\bvegafull \bdeltafull^\top + \bdeltafull \bvegafull^\top).
\end{equation}
The two-dimensional Kyle cross-impact model predicts that when trading options, one pushes the price in the amount of notional $\vega$ traded divided by the typical $\vega$ liquidity, which is compatible with findings from the meta-order study \cite{Toth2016TheMarkets}.
\\ \\
The full, four-dimensional Kyle cross-impact model $\lambdafourd$ (with calibrated prefactor $\prefactor = 0.5$) is shown in \cref{fig:kyle-4d}. Compared to the two-dimensional Kyle cross-impact model, it decouples the contribution of options on the level mode depending on the direction. This increases the explanatory power of the model, as will be clear in the next section.

\subsection{Explanatory power of cross-impact models}
\label{sec:quality_fit}

For practical applications, a good cross-impact model should explain realized price changes from order flows. Thus to compare the models previously introduced, we now examine their explanatory power on empirical data. Given a realization of the factor price process $(\delta \bp_t)_{1 \leq t \leq T}$ of length $T$, a corresponding series of predictions  $(\widehat{\delta \bp}_t)_{1 \leq t \leq T}$ and a symmetric positive semi-definite matrix $M$, we introduce the generalized $\rsqin(M)$ error as 
$$
\rsqin(M) := 1 - \dfrac{\sum_{1 \leq t \leq T} (\delta \bp_t - \widehat{\delta \bp}_t)^\top M (\delta \bp_t - \widehat{\delta \bp}_t)}{\sum_{1 \leq t \leq T} \delta \bp_t^\top M \delta \bp_t} \, .
$$
The matrix $M$ is used to examine a model's predictive power for different portfolios. As the factor of our system are natural directions to consider, we report $\rsqin(M)$ in \cref{table:scores} for $e_1 e_1^\top =: \Pi_{\textrm{spot}}$, $e_2 e_2^\top=: \Pi_{\textrm{level}}$, $e_3 e_3^\top =: \Pi_{\textrm{skew}}$ and $e_4 e_4^\top =: \Pi_{\textrm{term}}$.
\\
\begin{table}[h]
    \centering
    \scalebox{0.9}{\begin{tabular}{lcccccc} \toprule
    \multicolumn{1}{c}{Model} & \multicolumn{4}{c}{Scores} \\
    \cmidrule{2-5} & $\rsqin(\Pi_{\textrm{spot}})$ & $\rsqin(\Pi_{\textrm{level}})$ & $\rsqin(\Pi_{\textrm{skew}})$ & $\rsqin(\Pi_{\textrm{term}})$ \\
    \midrule
$\lambdabs$     & $ 0.18 \pm 0.01 $& $ -0.00 \pm 0.02 $& $ -0.00 \pm 0.01 $& $ -0.00 \pm 0.02 $ \\ 
$\lambdatwodnox$ & $ 0.18 \pm 0.01 $& $ -0.03 \pm 0.02 $& $ -0.01 \pm 0.01 $& $ 0.00 \pm 0.02 $ \\ 
$\lambdadirect$  & $ 0.18 \pm 0.01 $& $ -0.03 \pm 0.02 $& $ -0.14 \pm 0.02 $& $ -0.26 \pm 0.02 $ \\ 
$\lambdatwod$      & $ 0.20 \pm 0.01 $& $ 0.12 \pm 0.01 $& $ -0.01 \pm 0.01 $& $ 0.01 \pm 0.02 $ \\ 
$\lambdafourd$      & $ 0.20 \pm 0.01 $& $ 0.14 \pm 0.01 $& $ -0.12 \pm 0.02 $& $ 0.04 \pm 0.01 $ \\ 
    \bottomrule
    \end{tabular}}
    \caption{\textbf{Scores of different cross-impact models.\\}
    All scores were computed in-sample using the same data used for the calibration of the cross-impact models.}
    \label{table:scores}
\end{table}

\begin{figure}[h!]
	\centering
		\centering
		\includegraphics[width=0.8\linewidth]{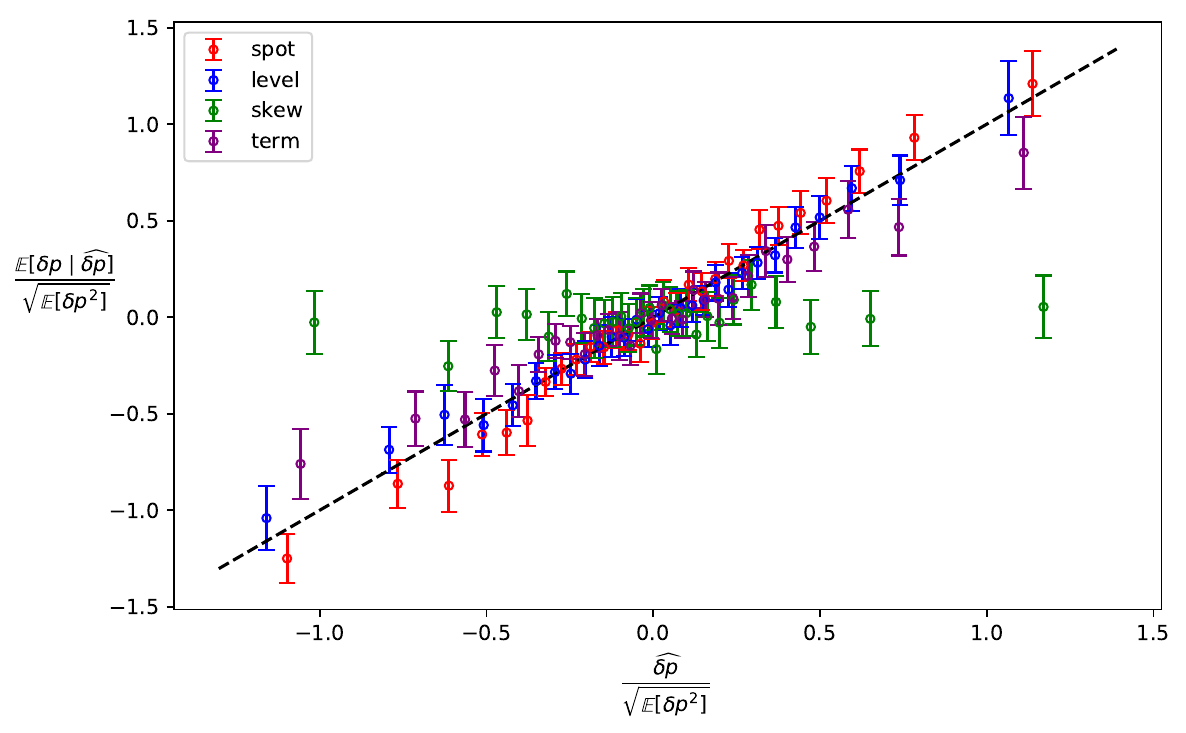}
		\caption{\textbf{Predictions of the four-dimensional Kyle model on the main directions of the system.} \\
		We report the expected price change conditional on the predicted price change of the four-dimensional Kyle model for the four main directions of the system: in red for the spot, blue for the level, green for the skew and purple for the term structure. Predicted price changes and conditional averages are both normalized by the standard deviation of price changes along the given direction.}
	    \label{fig:kyle_agg}
\end{figure}

The goodness-of-fit score in the spot direction $\rsqin(\Pi_{\textrm{spot}})$ is similar for all models, with cross-impact models being slightly better. Furthermore, there is no difference between $\lambdatwod$ and $\lambdafourd$. This is consistent with the liquidity reported in \cref{fig:risk_liquidity}. Indeed, most of the liquidity is placed on the spot and the order flow traded on other factors is small in comparison. Models which only take into account the spot thus capture most of the order flow explanatory power. There is also a small advantage in using order flow on the level mode since $\lambdatwod$ and $\lambdafourd$ score better, but using term and skew order flow provides no improvement.
\\ \\
While using solely spot liquidity to explain spot returns is a good approximation, $\rsqin(\Pi_{\textrm{level}})$ shows the same is not true for the level factor. Indeed, only models with cross-impact between spot and level factors properly explain changes in the level factor. This is natural as most of the traded order flow is on the spot but there is a high negative correlation between spot and level factor (see \cref{fig:risk_liquidity}). Unfortunately, all models fail to explain skew returns. We suspect this comes from the low signal to noise ratio and low liquidity (in risk terms) of the skew factor (see \cref{fig:risk_liquidity}). 
\\ \\
On all metrics, $\lambdafourd$ performs at least as well as $\lambdatwod$, which shows that the model is able to combine additional factors without suffering from noise. The additional factor also help weigh trades appropriately on the implied volatility surface, which improves the $\rsqin(\Pi_{\textrm{level}})$ score.  
\\ \\
Finally, we report the expected realized return conditional on the prediction of $\lambdafourd$ in \cref{fig:kyle_agg}. This shows that, skew aside, $\lambdafourd$ provides a good fit for the realized returns of the different factor as $\E[\delta \bp_t \mid \delta \hat{\bp}_t] \approx \delta \bp_t$. 

\subsection{Non-parametric evidence of cross-impact on options}
\label{sec:aggregate_impact}

\cref{sec:quality_fit} showed that only cross-impact models are able to explain returns for the level and term factor. Aside from this explanatory power, this section tests their ability to explain other features of our data. To do so, we introduce the cross aggregate impact metric. The cross aggregate impact induced from the portfolio $\bm{u} \in \reals{N+M}$ on the return of the portfolio $\bm{v} \in \reals{N+M}$ is
$$
\agg_{\bm{u},\bm{v}}(x) := \E [ \bm{v}^\top (\delta \bp_t, \delta \bP_t) \mid \bm{u}^\top (\delta \bq_t, \delta \bQ_t) = x] \, .
$$
If returns are given by a linear cross-impact model $\Psi$ and if we further assume $(\delta \bq_t, \delta \bQ_t)$ is a zero-mean Gaussian, then
\begin{align*}
    \agg_{\bm{u},\bm{v}}(x) &= \E [ \bm{v}^\top \Psi (\delta \bq_t, \delta \bQ_t) \mid \bm{u}^\top (\delta \bq_t, \delta \bQ_t)] := \agg_{\bm{u},\bm{v}}^{\Psi}(x) = a_{\Psi} x \, ,
\end{align*}
where the slope $a_{\Psi}$ depends on the cross-impact model $\Psi$ and on the order flow covariance. Even in the absence of cross-impact, the presence of order flow correlations between two portfolios $\bm{u}$ and $\bm{v}$ may lead to a non-zero cross aggregate impact. Thus, to test whether there is cross-impact, we compare the empirically measured $\agg_{\bm{u},\bm{v}}$ to the prediction $\agg_{\bm{u},\bm{v}}^{\Psi}$ for different cross-impact models $\Psi$. We differentiate models between those which have no off-diagonal contributions ($\lambdabs,\lambdadirect,\lambdatwodnox$) and thus ignore cross-impact and those that take it into account ($\lambdatwod,\lambdafourd$). 
\\ \\
We report $\agg_{\bm{u},\bm{v}}$ in \cref{fig:kyle_agg} for different portfolios $\bm{u},\bm{v}$ described in \cref{table:directions}. Diagonal plots show aggregate direct impact. As expected, buying the E-Mini increases, on average, the price of the E-Mini as shown by the $\bm{u}, \bm{v} = \text{spot}$ plot (first row, first column). We see from the $\bm{u},\bm{v} = \text{level}$ plot (second row, second column) that buying options and VIX futures increases, on average, the implied volatility. Furthermore, buying options and VIX futures decreases, on average, the E-Mini price as shown by the $\bm{u} = \text{level},  \bm{v}= \text{spot}$ plot (first row, second column). This same plot, among others of \cref{fig:kyle_agg}, shows that direct models provide a poor fit for cross aggregate impact. This suggests that the cross aggregate impact can only be explained by using a cross-impact model with off-diagonal elements, such as $\lambdafourd$. Further, the fit is noticeably better for $\lambdafourd$ than $\lambdatwod$ which highlights the importance of taking into account the skew and term factors.
\begin{table}[h!]
    \centering
    \scalebox{0.9}{\begin{tabular}{lcccccc} \toprule
    \multicolumn{1}{c}{Name} & \multicolumn{4}{c}{Components} \\
    \cmidrule{2-5} & spot & $\text{VIX}_0$ & $\text{VIX}_1$ & options \\
    \midrule
spot  & $(1,$ & $0,$ & $0,$ & $0, \cdots, 0)$ \\ 
level     & $(0,$ & $\beta^{11},$ & $\beta^{21},$ & $\beta^{31}, \cdots, \beta^{M1})$ \\ 
skew      & $(0,$ & $\beta^{12},$ & $\beta^{22},$ & $\beta^{32}, \cdots, \beta^{M2})$ \\ 
term & $(0,$ & $\beta^{13},$ & $\beta^{23},$ & $\beta^{33}, \cdots, \beta^{M3})$ \\ 
    \bottomrule
    \end{tabular}}
    \caption{\textbf{Description of different directions used in this section.}
    }
    \label{table:directions}
\end{table}

\begin{figure}[h!]
	\centering
		\centering
		\includegraphics[width=\linewidth]{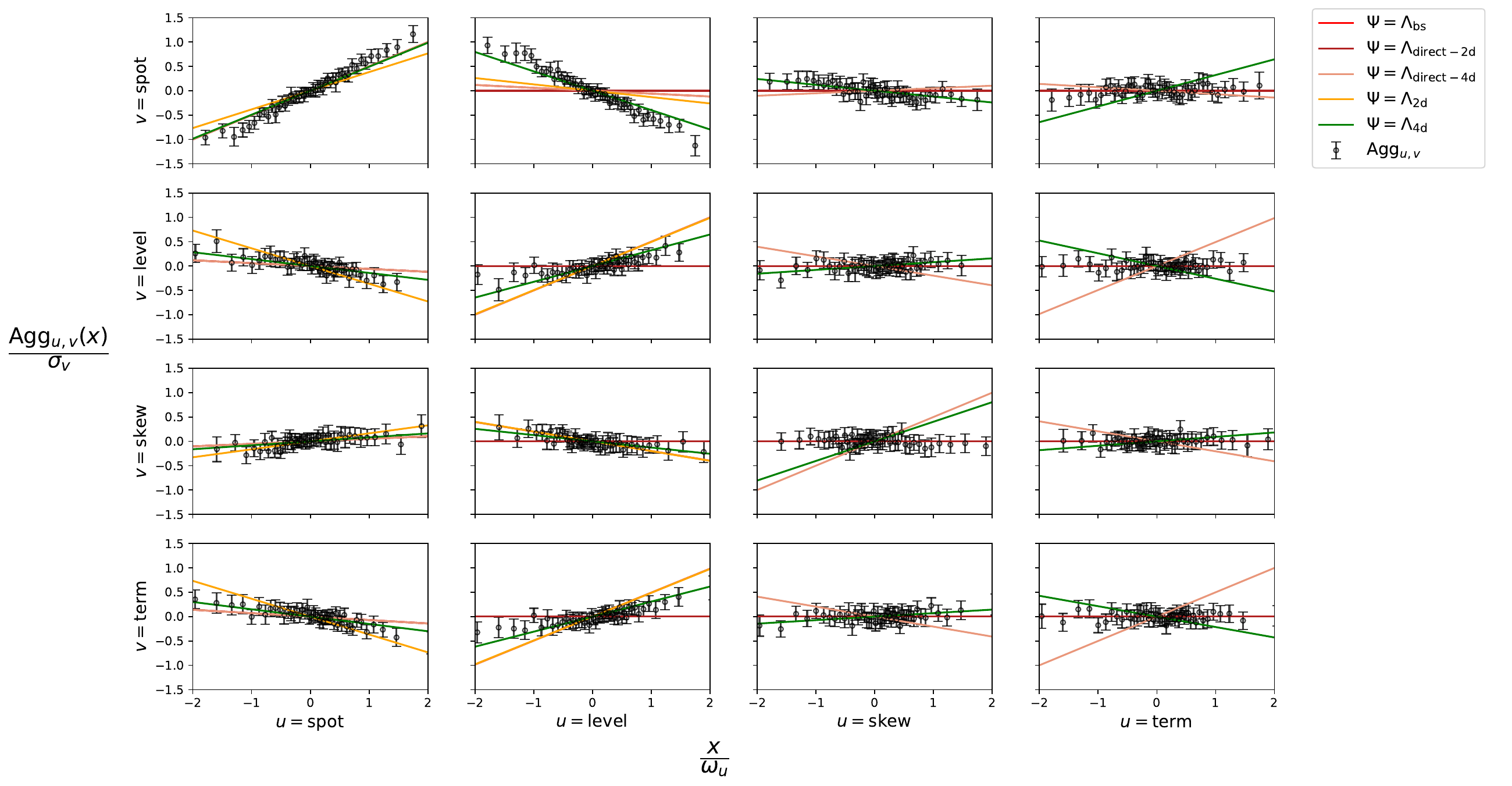}
		\caption{\textbf{Normalized cross aggregate impact curves.} \\
		We report the cross aggregate cross impact curves for the spot, level, skew and term structure directions. Aggregate traded volumes are normalized by the typical deviations $\omega_{u}^2 := \E_t[(\bm{u}^\top (\delta \bq_t, \delta \bQ_t))^2]$ and portfolio returns by the typical deviations $\sigma_{v}^2 := \E_t[(\bm{v}^\top (\delta \bp_t, \delta \bP_t))^2]$. Estimated cross aggregate impact $\agg_{\bm{u},\bm{v}}$ is reported along with predicted cross aggregate impact $\agg_{\bm{u},\bm{v}}^{\Psi}$ for different choices of linear cross-impact models $\Psi$.}
	    \label{fig:kyle_prefactor_agg}
\end{figure}

\section*{Conclusion}
\label{sec:conclusion}

The goal of this paper was to derive an estimation methodology for cross-impact on derivatives. To do so, we introduced a market where derivatives and factors, variables which determine the prices of derivatives, are co-traded and trades on one instrument induce impact on all instruments. We derived an explicit formula for cross-impact which is consistent with Ito's lemma and can be leveraged in practice. We applied the framework to E-Mini and VIX futures along with call and put options on the E-Mini.
\\ \\
One key result of this paper is an estimation formula for linear cross-impact which is adaptable to the derivative modeling framework and is compatible with popular stochastic volatility models. This formula fits in conveniently with existing option pricing frameworks and could be readily adapted in applications, provided one has access to the net traded order flow on derivatives.
\\ \\
Finally, our framework gives a recipe for aggregating liquidity of derivative markets which accounts for the joint dynamics of order flows and is tractable in practice. This is a topic of interest for regulators and practitioners alike, as option liquidity is very fragmented.

\bibliographystyle{plain}
\bibliography{bibliography_main.bib}

\clearpage

\appendix

\section{Proofs}
\label{app:proofs}

This section contains proofs of the results of \cref{sec:examples,sec:properties_impact}. We begin with the proof of \cref{prop:factoring_of}.

\begin{propositionp}{\ref*{prop:factoring_of}}
Using the notation of \cref{ass:impact_model,lemma:derivative_ito,ass:prices}, we have
\begin{align*}
\dd \bp_t &= \tbmup(\bp_t, t) \dd t + \frac{1}{\sqrt{1-\prefactor}} \volmatrixp(\bp_t, t) \dd B_t \\
\dd \bP_t &= \tbmuP(\bp_t, t) \dd t + \frac{1}{\sqrt{1-\prefactor}} \volmatrixP(\bp_t, t) \dd B_t \, ,
\end{align*}
where $\tbmup  \colon \reals{N} \times \reals{} \to \reals{N}$, $\tbmuP  \colon \reals{N} \times \reals{} \to \reals{M}$ and $B$ is a standard $N$ dimensional Brownian motion.
\end{propositionp}

\begin{proof}
We have, from \cref{eq:derivative_dynamics,eq:underlying_dynamics}
\begin{align*}
    \dd \bp_t &= \bmup(\bp_t, t) \dd t + \volmatrixp(\bp_t, t) \dd \bw_t + \cimpactpq(\bp_t, t) \dd \bq_t + \cimpactpQ(\bp_t, t) \dd \bQ_t \\
    \dd \bP_t &= \bmuP(\bp_t, t) \dd t + \volmatrixP(\bp_t, t) \dd \bw_t + \cimpactPq(\bp_t, t) \dd \bq_t + \cimpactPQ(\bp_t, t) \dd \bQ_t \, .
\end{align*}
Using the order flow dynamics given in \cref{ass:of_dynamics}, we have
\begin{align*}
    \dd \bp_t &= \left( \bmup(\bp_t, t) + \cimpactpq(\bp_t, t) \ofdrift_q(\bp_t, t)   + \cimpactpQ(\bp_t, t) \ofdrift_Q(\bp_t, t \right) \dd t \\
    & + \volmatrixp(\bp_t, t) \dd \bw_t + \cimpactpq(\bp_t, t) \ofdiffqq(\bp_t, t) \dd Z^q_t + \cimpactpQ(\bp_t, t) \ofdiffqQ(\bp_t, t) \dd Z^Q_t \\
    \dd \bP_t &= \left( \bmuP(\bp_t, t) + \cimpactPq(\bp_t, t) \ofdrift_q(\bp_t, t)   + \cimpactPQ(\bp_t, t) \ofdrift_Q(\bp_t, t \right) \dd t \\
    & + \volmatrixP(\bp_t, t) \dd \bw_t + \cimpactPq(\bp_t, t) \ofdiffQq(\bp_t, t) \dd Z^q_t + \cimpactPQ(\bp_t, t) \ofdiffQQ(\bp_t, t) \dd Z^Q_t \, .
\end{align*}
Thus, defining $\tbmup := \bmup + \cimpactpq \ofdrift_q   + \cimpactpQ \ofdrift_Q$ and $\tbmuP := \bmuP + \cimpactPq \ofdrift_q   + \cimpactPQ \ofdrift_Q$, the above reduces to
\begin{align*}
    \dd \bp_t &= \tbmup(\bp_t, t) \dd t + \volmatrixp(\bp_t, t) \dd \bw_t + \cimpactpq(\bp_t, t) \ofdiffqq(\bp_t, t) \dd Z^q_t + \cimpactpQ(\bp_t, t) \ofdiffqQ(\bp_t, t) \dd Z^Q_t \\
    \dd \bP_t &= \tbmuP(\bp_t, t) + \volmatrixP(\bp_t, t) \dd \bw_t + \cimpactPq(\bp_t, t) \ofdiffQq(\bp_t, t) \dd Z^q_t + \cimpactPQ(\bp_t, t) \ofdiffQQ(\bp_t, t) \dd Z^Q_t \, .
\end{align*}
The Kyle model is covariance-consistent (see \cite{tomas2020build} for a proof): we have $\cimpact(\bp_t, t) \Omega(\bp_t, t) \cimpact^\top(\bp_t, t) = Y \Sigma(\bp_t, t)$. By construction
$$
\covar{(\bp, \bP)} = \begin{pmatrix}
\volmatrixp \volmatrixp^\top +  & \volmatrixp \volmatrixP^\top \\
\volmatrixP \volmatrixp^\top & \volmatrixP \volmatrixP^\top
\end{pmatrix} + \cimpact \Omega \cimpact^\top = \Sigma
$$
So that we have 
$$
\begin{pmatrix}
\volmatrixp \volmatrixp^\top +  & \volmatrixp \volmatrixP^\top \\
\volmatrixP \volmatrixp^\top & \volmatrixP \volmatrixP^\top
\end{pmatrix} = (1-\prefactor) \Sigma
$$
Therefore, there exists a standard Brownian motion $B$ such that
\begin{align*}
    \dd \bp_t &= \tbmup(\bp_t, t) \dd t + \dfrac{1}{\sqrt{1-\prefactor}} \volmatrixp(\bp_t, t) \dd B_t \\
    \dd \bP_t &= \tbmuP(\bp_t, t) \dd t + \dfrac{1}{\sqrt{1-\prefactor}} \volmatrixP(\bp_t, t) \dd B_t \, .
\end{align*}
\end{proof}

We now prove \cref{prop:arbitrage}.

\begin{propositionp}{\ref*{prop:arbitrage}}
We have
\begin{equation*}
\Lambda(\bp_t, t) = \begin{pmatrix} \Lambda_{pq} & \Lambda_{pq} \Xi^\top \\ \Xi \Lambda_{pq} & \Xi \Lambda_{pq} \Xi^\top \end{pmatrix} (\bp_t, t) \, ,
\end{equation*}
where we recall that $\Xi := (\frac{\partial \bP^i}{\partial \bp^j})_{i,j}$ is the $M$ by $N$ sensitivity matrix.
\end{propositionp}

\begin{proof}

First, note that by \cref{ass:prices} combined with \cref{prop:factoring_of}, we have that $\bP_t = F(\bp_t, t)$ and
\begin{align*}
    \dd \bp_t &= \tbmup(\bp_t, t) \dd t + \dfrac{1}{\sqrt{1-\prefactor}} \volmatrixp(\bp_t, t) \dd B_t \\
    \dd \bP_t &= \tbmuP(\bp_t, t) \dd t + \dfrac{1}{\sqrt{1-\prefactor}} \volmatrixP(\bp_t, t) \dd B_t \, .
\end{align*}
Therefore, from using Ito's lemma, we have that $\volmatrixP = \Xi \volmatrixp$, where $\Xi_{ij} := \dfrac{\partial \bP^i}{\partial \bp^j}$. Therefore, as shown in the proof of \cref{prop:factoring_of}, we have
\begin{align*}
\begin{pmatrix}
\volmatrixp \volmatrixp^\top +  & \volmatrixp \volmatrixP^\top \\
\volmatrixP \volmatrixp^\top & \volmatrixP \volmatrixP^\top
\end{pmatrix} &= \begin{pmatrix}
\volmatrixp \volmatrixp^\top & \volmatrixp \volmatrixp^\top \Xi^\top \\
\Xi \volmatrixp \volmatrixp^\top & \Xi \volmatrixp \volmatrixp  \Xi^\top
\end{pmatrix} = (1-\prefactor) \Sigma
\end{align*}
Thus, the factor covariance matrix has the form
\begin{align*}
\Sigma &= \begin{pmatrix}
\Sigma_{pp} & \Sigma_{pp} \Xi^\top \\
\Xi \Sigma_{pp} & \Xi \Sigma_{pp}  \Xi^\top
\end{pmatrix} \, .
\end{align*}
The Kyle cross-impact model is fragmentation invariant (see \cite{tomas2020build} for a proof): if $x$ is in $\ker(\Sigma)$ then $x$ is  in $\ker(\cimpact)$. The above implies that, for every $u \in \R^M$, the vector $(\Xi^\top u, u)$ is in $\ker(\Sigma)$, so that, by fragmentation invariance,  $(\Xi^\top u, u)$ is in $\ker(\cimpact)$. Therefore, this implies that $\cimpactpQ = \cimpactpq \Xi^\top$. 
\\ \\
Since $\cimpact$ is symmetric, we also have that $\cimpactpq$ is symmetric and that $\cimpactPq = \cimpactpQ^\top =  \Xi \cimpactpq$. Another application of fragmentation invariance implies that $\cimpactPQ = \cimpactPq \Xi^\top = \Xi \cimpactpq \Xi^\top$. Summarising, we have
$$
\cimpact = \begin{pmatrix}
\cimpactpq & \cimpactpq \Xi^\top \\
\Xi \cimpactpq & \Xi \cimpactpq \Xi^\top
\end{pmatrix}
$$

\end{proof}

Finally, we prove the last proposition, \cref{prop:kyle_short}.

\begin{propositionp}{\ref*{prop:kyle_short}}
We have
\begin{equation*}
\Lambda_{pq} = \sqrt{\prefactor} (\Lc_{\Xi}^{-1})^\top \sqrt{\Lc_{\Xi}^\top \Sigma_{pp} \Lc_{\Xi}} \Lc_{\Xi}^{-1} \, ,
\end{equation*}
where we have omitted the dependence on $(\bp_t, t)$, $\OmegaXi := \Omegaqq + \Xi^\top \OmegaQQ \Xi + \Xi^\top \OmegaQq + \OmegaqQ \Xi$, and $\Lc_{\Xi}$ is a matrix such that $\Lc_\Xi \Lc_\Xi^\top = \OmegaXi$.
\end{propositionp}

\begin{proof}
From \cref{prop:arbitrage}, we have that 
$$
\cimpact = \begin{pmatrix}
\cimpactpq & \cimpactpq \Xi^\top \\
\Xi \cimpactpq & \Xi \cimpactpq \Xi^\top
\end{pmatrix}
$$
The Kyle cross-impact model is covariance-consistent (see \cite{tomas2020build} for a proof), i.e. we have
$$
\begin{pmatrix}
\Sigma_{pp} & \Sigma_{pP} \\
\Sigma_{Pp} & \Sigma_{PP}
\end{pmatrix} = Y \cimpact \Omega \cimpact^\top.
$$
Therefore, we obtain from the above that
$$
\Sigmapp = \prefactor \cimpactpq \OmegaXi \cimpactpq^\top, 
$$
where $\OmegaXi := \Omegaqq + \Xi^\top \OmegaQQ \Xi + \Xi^\top \OmegaQq + \OmegaqQ \Xi$. In particular, note that $\OmegaXi$ is a symmetric positive definite matrix. Since $\cimpact$ is a symmetric positive definite matrix (see \cite{tomas2020build} for a proof), so is $\cimpactpq$. From \cite{tomas2020build}, we know that $\cimpactpq$ is the unique symmetric positive definite solution of the system $\Sigmapp = \cimpactpq \OmegaXi \cimpactpq^\top$ and that its form is given by
$$
\cimpactpq = \sqrt{\prefactor} (\OmegaXi^{-1/2})^\top \sqrt{(\OmegaXi^{1/2})^\top \Sigmapp \OmegaXi^{1/2}} \OmegaXi^{-1/2} \, .
$$
\end{proof}

\begin{lemmap}{\ref*{lemma:kyle_leverage}}
We denote by $\bdeltafull := (1, 0, \frac{\partial P^1}{\partial s}, \cdots, \frac{\partial P^M}{\partial s})$, $\bvegafull := (0, 1, \frac{\partial P^1}{\partial \varsigma}, \cdots, \frac{\partial P^M}{\partial \varsigma})$ the vectors of sensitivities. Then, if $\bdeltafull^\top \Omega \bvegafull = 0$, we have
\begin{equation*}
\cimpactpq(s_t, \varsigma_t, t) = \dfrac{\sqrt{\prefactor}}{\sqrt{\sigma^2 \omega_{\Delta}^2 + \xi^2 \omega_{\vega}^2 + 2 \sigma \xi \rho \omega_{\Delta} \omega_{\vega}}}
\begin{pmatrix}
\sigma^2 + \frac{\omega_{\vega}}{\omega_{\Delta}} \sigma \xi \sqrt{1-\rho^2} & \sigma \xi \rho \\
\sigma \xi \rho & \xi^2 + \frac{\omega_{\Delta}}{\omega_{\vega}} \sigma \xi \sqrt{1-\rho^2}
\end{pmatrix} (s_t, \varsigma_t, t) \, ,
\end{equation*}
where $\omega_{\Delta}^2 := \bdeltafull^\top \Omega \bdeltafull$ is the delta-aggregated liquidity, $\omega_{\vega}^2 := \bvegafull^\top \Omega \bvegafull$ is the vega-aggregated liquidity, $\sigma^2(s_t, \varsigma_t, t) \dd t := \dd \covar{s,s}_t$ is the spot volatility, $\xi^2(s_t, \varsigma_t, t) \dd t:= \dd \covar{\varsigma, \varsigma}_t$ is the volatility of volatility and $\rho(s_t, \varsigma_t, t) \dd t := \dfrac{\dd \covar{\varsigma, s}_t}{\xi(s_t, \varsigma_t, t) \sigma(s_t, \varsigma_t, t)}$ is the spot-vol correlation.
\end{lemmap}

\begin{proof}
Using the results of \cref{prop:kyle_short}, the cross-impact matrix $\cimpactpq$ is of the form
$$
\cimpactpq(s_t, \varsigma_t, t) \sqrt{\prefactor} (\Lc_{\Xi}^{-1})^\top \sqrt{\Lc_{\Xi}^\top \Sigma_{pp} \Lc_{\Xi}} \Lc_{\Xi}^{-1} (s_t, \varsigma_t, t) \, ,
$$
where $\OmegaXi := \Omegaqq + \Xi^\top \OmegaQQ \Xi + \Xi^\top \OmegaQq + \OmegaqQ \Xi$, and $\Lc_{\Xi}$ is a matrix such that $\Lc_\Xi \Lc_\Xi^\top = \OmegaXi$. Using the notations of the lemma, we have
$$
\Sigma_{pp}(s_t, \varsigma_t, t) = \begin{pmatrix}
\sigma^2 & \sigma \xi \rho \\
\sigma \xi \rho & \xi^2
\end{pmatrix}(s_t, \varsigma_t, t) \hspace{1cm} \OmegaXi(s_t, \varsigma_t, t) = \begin{pmatrix}
\omega_{\Delta}^2 & 0 \\
0 & \omega_{\vega}^2
\end{pmatrix}(s_t, \varsigma_t, t) \, .
$$
Plugging the above into the formula for $\cimpactpq$ and applying some straightforward linear algebra gives the result. 
\end{proof}

\section{Empirical details}

\subsection{Data}
\label{app:data}

This section gives motivation about our choice of instruments, details on the data and methodology which were omitted in the main text for conciseness.

\begin{figure}[h!]
    \centering
        \centering
   \hspace{1.2cm} \includegraphics[width=0.6\linewidth]{figures/risk_liquidity.pdf}
    \caption{\textbf{Distribution of liquidity among VIX futures and options.}\\}
    \label{fig:risk_liquidity}
\end{figure}

\paragraph{Choice of instruments} To stress-test our approach, we sought an actively traded derivative market with many derivatives. Thus we considered E-mini vanilla options and their factor (both quoted on the CME), the front-month futures contract. However, a large fraction of the traded risk in derivatives comes from VIX futures (quoted on the CBOE) as shown in \cref{fig:risk_liquidity}. The VIX index is computed using options with maturities between 23 and 37 days and is meant to track the level of the implied volatility for options expiring in one month. Thus, because of their liquidity and close relationship with the implied volatility of options, order flow traded on VIX futures play an important role and should not be ignored.

\paragraph{Filtering instruments} Given the very large number of options quoted on the market, we kept options within a given range of strikes and maturities to limit the size of the data set.

\paragraph{Resolution and time frame}
Our dataset contains the trades and quotes of all previously selected products, at the five minute time scale, from January 2019 to September 2019. This time frame was chosen because of the large level of noise on derivatives' prices and the size of the data set which encumbered analysis. In a given five minute bin, signed trades were aggregated on their volumes, so that we have the opening and closing prices of instruments along with the aggregated signed traded order flow. We considered hours where both options and their factor are liquid, further removing 30 minutes around opening and closing for stationarity purposes. Doing so, data ranges between 3PM and 8:30PM UTC.

\paragraph{Implied volatility and greeks}
We now explain how implied volatility and Greeks were computed. For a given day and for a particular option, we have access to the opening bid and ask prices of that option for each five minute window. Furthermore, the bid and ask Black-Scholes implied volatilities are computed using the bid and ask price of the option and the price of the E-mini future contract with closest maturity. Correspondingly, the usual Black-Scholes Greeks $\Delta$ and $\vega$ are computed for both the bid and ask sides. In our analysis, we use the mid of bid and ask quantities (option prices, implied volatilities and greeks) to perform computations.

\subsection{Estimation of covariances}
\label{app:covariances}

This section details the estimation of the covariances $\Sigma_{pp}$ and $\OmegaXi$ used in \cref{sec:empirical}.

\paragraph{Estimation of $\OmegaXi$}
In order to compute $\OmegaXi$, we begin by computing $\Xi(\bp_t, t)$. Using \cref{prop:factoring_of}, the sensitivities of the option prices are computed according to the usual Greeks adjusted for the option's sensitivity as described in \cref{sec:vol_factors_example}. Thus, for a given time window $[t, t + \Delta t]$ of length $\Delta t = \text{5 minutes}$, we measure the net traded order flow on factors $\delta \bq_t$ and on derivatives $\delta \bQ_t$. We weigh order flow traded on derivatives according to sensitivities and build the aggregate net order flow on that time window: $\delta \bq_t + \Xi(\bp_t, t) \delta \bQ_t$. We measure this quantity over all available time windows and find that it is roughly stationary and does not depend on the value of the factors. This motivates dropping the dependence of these variables and the estimation of $\OmegaXi$ as the covariance of the aggregate order flow: $\OmegaXi \approx \cov(\delta \bq_t + \Xi(\bp_t, t) \delta \bQ_t)$.

\paragraph{Estimation of $\Sigma_{pp}$}
After selection of the relevant factors detailed in \cref{sec:setupemp}, we estimate the factor return covariance matrix in the following manner. Given a time window $[t, t + \Delta t]$ of length $\Delta t = \text{5 minutes}$, we measure the opening and closing price of all factors. The spot factor's price is taken as the price of the E-Mini future closest to expiry. For options, the factor values are computed by projecting option implied volatilities in the direction of the implied volatility surfaces: $\varsigma^q = \sum_{i=1}^{M} \beta^{iq} \hat{\sigma}^i$. This formula holds since, by construction, $(\beta^{iq})$ satisfies $\beta^\top \beta = \id$. Thus, we compute opening and closing prices of the factors of the implied volatility surface. The factor covariance matrix $\Sigma_{pp}$ is then estimated as the covariance matrix of price changes.

\end{document}